\documentclass[aps,prl,twocolumn,superscriptaddress,calc,epsfig]{revtex4-1}
\usepackage{graphicx}
\usepackage{amsfonts}
\usepackage{amsmath}
\usepackage{amssymb}
\usepackage{epsfig}

\begin{document}

\title{Evolution of Fermion Pairing from Three to Two Dimensions}

\author{Ariel T. Sommer}
\author{Lawrence W. Cheuk}
\author{Mark J. H. Ku}
\author{Waseem S. Bakr}
\author{Martin W. Zwierlein}
\affiliation{Department of Physics, MIT-Harvard Center for Ultracold Atoms, and Research Laboratory of Electronics,
        MIT, Cambridge, Massachusetts 02139, USA}

\newcommand{\li}[1]{\ensuremath{^{#1}\mathrm{Li}}}
\newcommand{\na}[1]{\ensuremath{^{#1}\mathrm{Na}}}
\newcommand{\ket}[1]{\ensuremath{|#1\rangle}}
\newcommand{\eb}{\ensuremath{E_{\mathrm{b}}}}
\newcommand{\ebtheory}{\ensuremath{E_{\mathrm{b,2-body}}}}
\newcommand{\ebmanybody}{\ensuremath{E_{\mathrm{b}}}}
\newcommand{\ebmeanfield}{\ensuremath{E_{\mathrm{b,MF}}}}
\newcommand{\ebtwobody}{\ensuremath{E_{\mathrm{b}}}}
\newcommand{\ebf}{\ensuremath{E_{\mathrm{b}}'}}
\newcommand{\nubb}{\ensuremath{\nu_{\mathrm{bb}}}}
\newcommand{\depth}{\ensuremath{V_0}}
\newcommand{\latfreq}{\ensuremath{\omega_z}}
\newcommand{\er}{\ensuremath{E_R}}
\newcommand{\wm}{\ensuremath{w_m}}
\newcommand{\eftwo}{\ensuremath{E_F^{2\mathrm{D}}}}
\newcommand{\omegarf}{\ensuremath{\omega_{\mathrm{rf}}}}
\newcommand{\omegahf}{\ensuremath{\omega_{\mathrm{hf}}}}
\newcommand{\nurf}{\ensuremath{\nu_{\mathrm{rf}}}}
\newcommand{\nuhf}{\ensuremath{\nu_{\mathrm{hf}}}}
\newcommand{\widthm}{\ensuremath{w_m}}
\newcommand{\widtha}{\ensuremath{w_a}}
\newcommand{\atwod}{\ensuremath{a_{2\mathrm{D}}}}
\newcommand{\lnkfa}{\ensuremath{\mathrm{ln}(k_F\atwod)}}

\newcommand{\orso}{Orso \textit{et al.}~\cite{orso05form}}

\begin{abstract}
We follow the evolution of fermion pairing in the dimensional crossover from 3D to 2D as a strongly interacting Fermi gas of $^6$Li atoms becomes confined to a stack of two-dimensional layers formed by a one-dimensional optical lattice. Decreasing the dimensionality leads to the opening of a gap in radio-frequency spectra, even on the BCS-side of a Feshbach resonance. The measured binding energy of fermion pairs closely follows the theoretical two-body binding energy and, in the 2D limit, the zero-temperature mean-field BEC-BCS theory.
\end{abstract}
\maketitle
Interacting fermions in coupled two-dimensional (2D) layers present unique physical phenomena and are central to the description of unconventional superconductivity in high-transition-temperature cuprates~\cite{tink04scLD} and layered organic conductors~\cite{lang96quasi}. Experiments on ultracold gases of fermionic atoms have allowed access to the crossover from Bose-Einstein condensation (BEC) of tightly-bound fermion pairs to Bardeen-Cooper-Schrieffer (BCS) superfluidity of long-range Cooper pairs in three spatial dimensions~\cite{ingu08varenna,giorgini08theory}, and more recently, the confinement of interacting Fermi gases to two spatial dimensions~\cite{gunter05pwave,du09inelastic,martiyanov10observation,frohlich11radio,dyke11crossover}. A fermionic superfluid loaded into a periodic potential should form stacks of two-dimensional superfluids with tunable interlayer coupling~\cite{orso05super,orso05form,salasnich07condensate,zhang08bcs}, an ideal model for Josephson-coupled quasi-2D superconductors~\cite{rugg80quasi2D,tink04scLD}. For deep potentials in the regime of uncoupled 2D layers, increasing the temperature of the gas is expected to destroy superfluidity through the Berezinskii-Kosterlitz-Thouless mechanism~\cite{kosterlitz72long,petrov03superfluid,zhang08berez}, while more exotic multi-plane vortex loop excitations are predicted for a 3D-anisotropic BCS superfluid near the critical point~\cite{iskin09evolution}.

\begin{figure}
    \centering
    \includegraphics[width=3in]{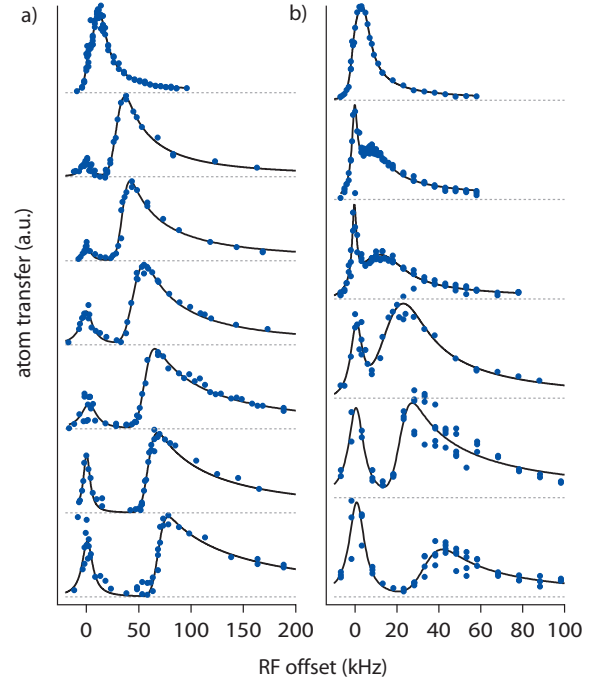}
    \caption{\label{fig:spec}(Color online) Evolution of fermion pairing in the 3D to 2D crossover in a one-dimensional optical lattice, observed via RF spectroscopy. Shown is the transferred atom number versus RF offset frequency relative to the atomic hyperfine splitting. a) Spectra at the Feshbach resonance at 690.7(1) G with $d/a=-0.01(4)$. Lattice depths from top to bottom in units of \er: 1.84(3), 4.8(2), 6.1(2), 9.9(4), 12.2(4), 18.6(7), and 19.5(7). b) Spectra on the BCS-side at 720.7(1) G, $d/a=-1.15(2)$. Lattice depths in units of \er: 2.75(5), 4.13(7), 4.8(1), 6.0(2), 10.3(2) 18.1(4).}
\end{figure}

In this work, we study fermion pairing across the crossover from 3D to 2D in a periodic potential of increasing depth.
To form a bound state in 3D, the attraction between two particles in vacuum must exceed a certain threshold.
However, if the two particles interact in the presence of a Fermi sea, the Cooper mechanism allows pairing for arbitrarily weak interactions~\cite{coop56boun}. In 2D, even two particles in vacuum can bind for arbitrarily weak interactions. Surprisingly, the mean-field theory of the BEC-BCS crossover in 2D predicts that the binding energy of fermion pairs in the many-body system is identical to the two-body binding energy \ebtwobody~\cite{rand89}. Indeed, to break a pair and remove one pairing partner from the system costs an energy~\cite{kett08maki} $\ebmeanfield= \sqrt{\mu^2+\Delta^2} - \mu$ within mean-field theory, where $\mu$ is the chemical potential and $\Delta$ is the pairing gap. In 2D, one finds~\cite{rand89} $\mu = E_F - \ebtwobody/2$ and $\Delta^2 = 2 E_F \ebtwobody$, where $E_F$ is the Fermi energy, and thus $\ebmeanfield= \ebtwobody$, i.e. the many-body and two-body binding energies are predicted to be identical throughout the BEC-BCS crossover.

We realize a system that is tunable from 3D to 2D with a gas of ultracold fermionic \li{6} atoms trapped in an optical trap and a standing-wave optical lattice. The lattice produces a periodic potential along the $z$ direction,
\begin{equation}
    V(z) = V_0\, \mathrm{sin}^2(\pi z/d),
\end{equation}
with depth $V_0$ and lattice spacing $d=532$ nm. Together with the optical trap, the lattice interpolates between the 3D and 2D limits. It gradually freezes out motion along one dimension and confines particles in increasingly uncoupled layers. Features characteristic of the 2D system appear as the strength of the periodic potential is increased. The threshold for pairing is reduced, allowing pairs to form for weaker attractive interactions than in the 3D system. The effective mass of particles increases along the confined direction, and center-of-mass and relative degrees of freedom of an atom pair become coupled~\cite{orso05form}. For a deep potential that suppresses interlayer tunneling, the system is an array of uncoupled two-dimensional layers. Here, center of mass and relative motion decouple and fermion pairs form for the weakest interatomic attraction~\cite{orso05form,petr01inte,bloc08many}.

In the experiment, the appearance of bound fermion pairs is revealed using radio-frequency (RF) spectroscopy. The atomic gas consists of an equal mixture of \li{6} atoms in the first and third hyperfine states (denoted \ket{1} and \ket{3}), chosen to minimize final state interaction effects in the RF spectra~\cite{schu08dete}. Interactions between atoms in state \ket{1} and \ket{3} are greatly enhanced by a broad Feshbach resonance at 690.4(5) G~\cite{bart05prec}. An RF pulse is applied to transfer atoms from one of the initial hyperfine states to the unoccupied second hyperfine state (denoted \ket{2}). In previous work on RF spectroscopy of $^{40}$K fermions in a deep 1D lattice~\cite{frohlich11radio}, an RF pulse transferred atoms from an initially weakly interacting state into a strongly interacting spin state, likely producing polarons~\cite{pietila11pairing}. In our work the initial state is the strongly interacting, largely paired Fermi gas in equilibrium, and the final state is weakly interacting.

An asymmetric dissociation peak (the bound to free transition) in the RF spectrum indicates the presence of fermion pairs.  For two-particle binding, the pair dissociation lineshape in the 3D and 2D limits is proportional to $\rho(h\nu-\eb)/\nu^2$, with $\rho$ the free-particle density of states, and $\nu=\pm(\nurf-\nuhf)$ the offset of the RF frequency $\nurf$ from the hyperfine splitting \nuhf{} (plus symbol: $\ket{1}\rightarrow\ket{2}$ transition, minus symbol: $\ket{3}\rightarrow\ket{2}$ transition). This form can be obtained from Fermi's Golden Rule and the bound state wavefunction in momentum space; see also Refs.~\cite{chin05radi,kett08maki}. In 2D, the expected dissociation lineshape is then proportional to
\begin{equation}
    I(\nu) \propto \frac{\theta(h\nu-\eb)}{\nu^2}.
    \label{eqn:molec}
\end{equation}

In addition to the pairing peak, at finite temperature one expects a peak in the RF spectrum due to unbound atoms, the free to free transition.
A narrow bound to bound transition can also be driven at an offset frequency $\nubb=(\eb-\ebf)/h$ that transfers one spin state of the initial bound pair with binding energy $\eb$ into a bound state of \ket{2} with \ket{1} or \ket{3}, of binding energy \ebf. For a \ket{1}-\ket{3} mixture near the Feshbach resonance, $\eb\ll\ebf$~\cite{schu08dete}, so the bound to bound peak is well separated from the bound to free and free to free peaks. As very recently calculated~\cite{lang11clock2d}, final state interactions and the anomalous nature of scattering in 2D introduce an additional factor of $\frac{\ln^2(\eb/\ebf)}{\ln^2((h\nu-\eb)/\ebf)+\pi^2}$ into Eq.~\ref{eqn:molec}, causing a rounding off of the sharp peak expected from the step-function.

In a 1D lattice, the binding energy for two-body pairs is determined by the lattice spacing $d$, the depth $V_0$, and the 3D scattering length $a$. In the 2D limit $\depth\gg\er$, with recoil energy ${\er=\frac{\hbar^2\pi^2}{2md^2}}$, the scattering properties of the gas are completely determined by $\eb$~\cite{bloc08many,petr01inte}. In that limit, the lattice wells can be approximated as harmonic traps with level spacing ${\hbar\omega_z=2\sqrt{\depth\er}}$ and harmonic oscillator length ${l_z=\sqrt{\frac{\hbar}{m\omega_z}}}$. In a many-particle system in 2D, the ratio of the binding energy to the Fermi energy determines the strength of interactions. The 2D scattering amplitude ${f(E_F)=\frac{2\pi}{-\lnkfa+i\pi/2}}$ for collisions with energy $E_F$ is parameterized by \lnkfa, where ${k_F=\sqrt{2mE_F}/\hbar}$ and $\atwod=\hbar/\sqrt{m\eb}$. It is large when ${|\lnkfa|\lesssim 1}$~\cite{petr01inte,bloc08many}, corresponding to the strong-coupling regime~\cite{bert11crossover, lang11clock2d}. The BEC side of the BEC-BCS crossover corresponds to negative values of $\lnkfa$, while the BCS side corresponds to positive values~\cite{rand89}.


The experimental sequence proceeds as follows. An ultracold gas of \li{6} is produced by sympathetic cooling with \na{23} as described previously~\cite{kett08maki}. The \li{6} atoms are transferred from a magnetic trap to an optical dipole trap (wavelength 1064 nm, waist 120 $\mu$m), with axial harmonic confinement (frequency 22.8 Hz) provided by magnetic field curvature. With \li{6} polarized in state \ket{1}, the magnetic bias field is raised to 568 G, and an equal mixture of hyperfine states \ket{1} and \ket{3} is created using a 50\% RF transfer from \ket{1} to \ket{2} followed by a full transfer from \ket{2} to \ket{3}. The field is then raised to the final value and evaporative cooling is applied by lowering the depth of the optical dipole trap, resulting in a fermion pair condensate with typically $5\times10^5$ atoms per spin state. The lattice is then ramped up over 100 ms. The retro-reflected lattice beam (wavelength 1064 nm) is at an angle of 0.5 degrees from the optical dipole trap beam, enough to selectively reflect only the lattice beam. The depth of the lattice is calibrated using Kapitza-Dirac diffraction of a \na{23} BEC and a $\li{6}_2$ molecular BEC, and by lattice modulation spectroscopy on the \li{6} cloud.  The magnetic field and hyperfine splitting are calibrated using RF spectroscopy on spin polarized clouds. After loading the lattice, the RF pulse is applied for a duration of typically 1 ms. Images of state \ket{2} and either \ket{1} or \ket{3} are recorded in each run of the experiment.

To ensure loading into the first Bloch band, the Fermi energy and temperature of the cloud are kept below the second band. The 2D Fermi energy $\eftwo=\frac{2\pi\hbar^2n}{m}$, with $n$ the 2D density per spin state, is typically $h\cdot10$ kHz. The bottom of the second band is at least one recoil energy $\er=h\cdot29.3$ kHz above the bottom of the first band in shallow lattices, and up to about $h\cdot300$ kHz for the deepest lattices. The temperature is estimated to be on the order of the Fermi energy.

RF spectra are recorded for various lattice depths and interaction strengths. Figure~\ref{fig:spec} shows examples of spectra over a range of lattice depths at the 3D Feshbach resonance and on the BCS-side of the resonance at 721 G, where fermion pairing in 3D is a purely many-body effect. At the lowest lattice depths, the spectra show only a single peak, shifted to positive offset frequencies due to many-body interactions. This is similar to the case without a lattice ~\cite{schu08dete,schi08dete}; to discern a peak due to fermion pairs from a peak due to unbound atoms would require locally-resolved RF spectroscopy of imbalanced Fermi gases~\cite{schi08dete}. However, as the lattice depth is raised, the single peak splits into two and a clear pairing gap emerges. The narrow peak at zero offset is the free to free transition, and the asymmetric peak at positive offset is the pair dissociation spectrum. The pair spectrum, especially on resonance, shows a sharp threshold, and a long tail corresponding to dissociation of fermion pairs into free atoms with non-zero kinetic energy.

\begin{figure}
    \centering
    \includegraphics{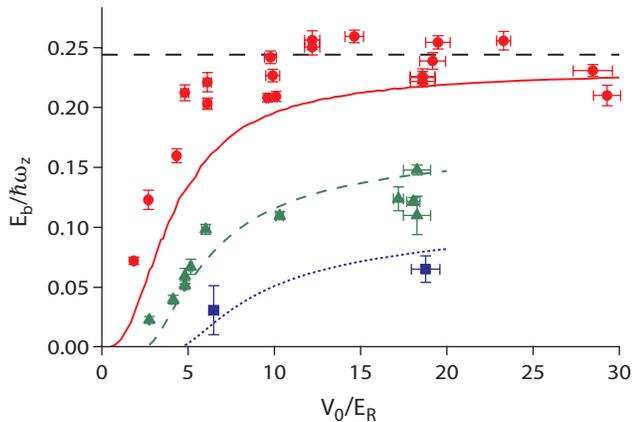}
    \caption{\label{fig:eb}(Color online) Binding energy \eb{} versus lattice depth \depth{} at several values of the 3D scattering length $a$. \eb{} is normalized via the lattice frequency $\omega_z$. Red circles: results from spectra at 690.7(1) G and $d/a=-0.01(4)$. Green triangles: 720.7(1) G, $d/a=-1.15(2)$. Blue squares: 800.1(1) G, $d/a=-2.69(1)$. Curves show predictions from \orso. Black dashed line: harmonic approximation result for $1/a=0$.}
\end{figure}

Binding energies are determined from the offset frequency of the pairing threshold. Although the lineshape in Eq.~(\ref{eqn:molec}) jumps discontinuously from zero to its maximum value, the spectra are observed to be broadened. This is to a large part due to the logarithmic corrections~\cite{lang11clock2d} noted above, that predict a gradual rise at the threshold $h\nu=\eb$, and a spectral peak that is slightly shifted from $\eb$. We include possible additional broadening by convolving the theoretical lineshape, including the logarithmic correction, with a gaussian function of width $\widthm$. The parameters $\eb$ and $\widthm$ are determined by a least-squares fit to the measured spectrum. Typical spectra have $\widthm$ of 5 kHz, consistent with our estimates of broadening based on collisions and three-body losses. The Fourier broadening is 1 kHz. Power broadening is about 5 kHz on the free to free transition, and less than 1 kHz on the bound to free transition due to the reduced wavefunction overlap. Inclusion of the logarithmic correction is found to be necessary in order for the fit function to reproduce the observed behavior of the high frequency tail. The final state binding energy used in the logarithmic correction for fitting is obtained from spectra where both a bound to bound and a bound to free peak were measured. At low lattice depths, the 2D form for the paired spectrum should differ from the exact shape that interpolates between the 3D and 2D limits. In the case where the shape of the spectrum is given by the 3D limit, fitting to the 2D form overestimates the binding energy by 8\%.

\begin{figure}
    \centering
    \includegraphics{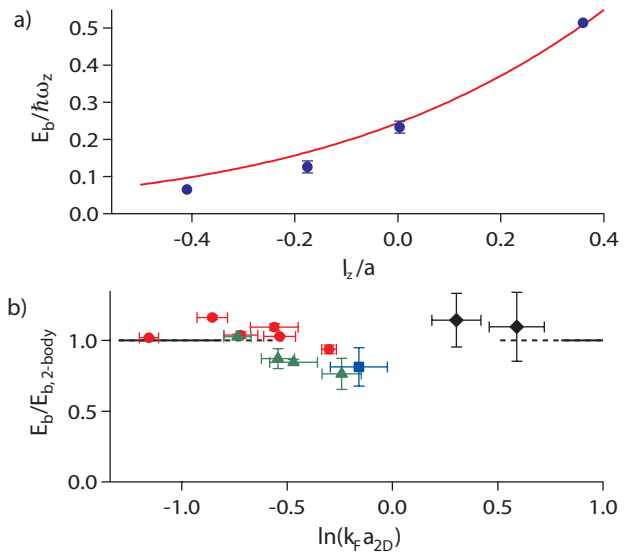}
    \caption{\label{fig:ebtwo}(Color online) a) Binding energy of fermion pairs versus interaction strength $l_z/a$ for deep lattices ($\depth > 17 \er$). Solid curve: theoretical prediction in the 2D harmonic limit~\cite{petr01inte,bloc08many}. b) Ratio of the measured binding energy to the two-body result~\cite{orso05form} versus \lnkfa{} for $\depth > 17 \er$. Black diamonds: binding energy determined from the bound to bound transition with resonant final state interactions. Other data symbols: see Fig.~\ref{fig:eb}. Horizontal line: zero-temperature mean-field theory~\cite{rand89}.}
\end{figure}

Figure~\ref{fig:eb} shows the measured binding energies as function of $\depth/\er$ for several interaction strengths. The binding energies are normalized by $\hbar\omega_z \equiv 2\sqrt{\depth\er}$, which equals the level spacing in the harmonic approximation to the lattice potential.
The measured binding energies grow with increasing lattice depth, and agree reasonably well with theoretical predictions for two-body bound pairs in a 1D lattice~\cite{orso05form}. The binding energy at the 3D resonance approaches a constant multiple of $\hbar\omega_z$ as the lattice depth increases, as expected from the 2D limit~\cite{petr01inte,bloc08many}. Figure~\ref{fig:ebtwo}(a) compares the binding energies measured in lattices deeper than $17\er$ to predictions in the harmonic quasi-2D limit~\cite{petr01inte,bloc08many}. At the 3D Feshbach resonance, we find $\eb=0.232(16)\hbar\omega_z$ for deep lattices. The error bar refers to the standard error on the mean. This value is close to the harmonic confinement result of $0.244\hbar\omega_z$~\cite{bloc08many}. The exact calculation~\cite{orso05form} predicts a constant downward shift of the binding energy by 0.2\er{} for deep lattices due to the anharmonicity of the sinusoidal potential. For \depth{} of about 20\er{}, this gives a prediction of $0.22\hbar\omega_z$, also close to the measured value.

Figure~\ref{fig:ebtwo}(b) shows the binding energy measured in deep lattices normalized by the exact two-body result~\cite{orso05form} versus the many-body interaction parameter $\lnkfa$.
Overall, the binding energies are close to the two-body value, even in the strong coupling regime for $|\lnkfa|<1$, as predicted by zero-temperature mean-field theory~\cite{rand89}. The data show a slight downward deviation for the strongest coupling.
At fixed reduced temperature $T/T_F$, the relationship should be universal. It will thus be interesting to see in future work whether the binding energy depends significantly on temperature.

\begin{figure}
    \centering
    \includegraphics[width=3in]{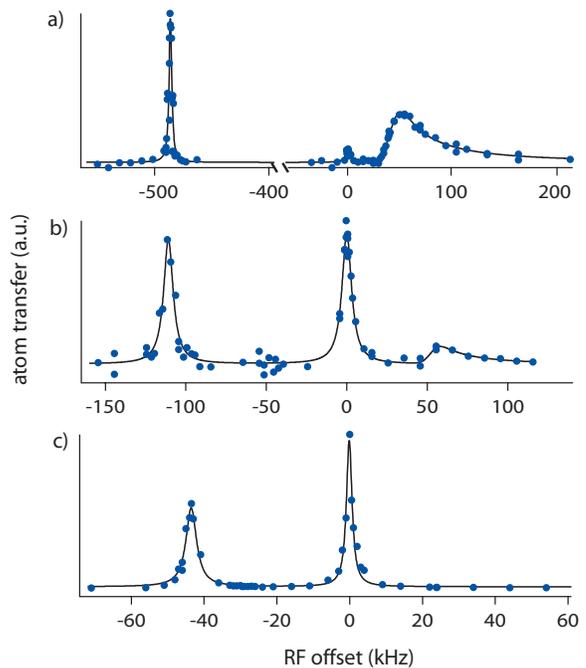}
    \caption{\label{fig:bb} (Color online) Spectra including the bound to bound transition, a narrow peak at negative RF offset. Shown are spectra at magnetic fields of a) 690.7(1) G, b) 751.1(1) G and c) 834.4(1). The interaction parameters $d/a$ are a) -0.01(4), b) -1.91(1), and c) -3.06(1). Lattice depths in units of \er{} are a) 9.59(7), b) 26.1(4), and c) 26.4(3). The bound to free transition is not visible in (c). The transfer is from \ket{1} to \ket{2} in (a) and (b) and from \ket{3} to \ket{2} in (c).}
\end{figure}

The bound to bound transition is seen in Fig.~\ref{fig:bb} as a narrow peak at negative offset frequencies. In the regime where \eb{} can be found from the pair dissociation spectrum, the bound to bound peak position directly yields the binding energy in the final state \ebf. For example, the spectrum in Fig.~\ref{fig:bb}(a), taken at the 3D \ket{1}-\ket{3} resonance at 690.7(1) G and $\depth/\er=9.59(7)$, gives $\ebf/\er=18.0(1)$ at a final state interaction of $d/a'=8.41(2)$. Likewise, the spectrum in Fig.~\ref{fig:bb}(b) at $\depth/\er=26.1(4)$ and a magnetic field of 751.1(1) G, where $d/a'=2.55(1)$, gives $\ebf/\er=5.3(1)$. An independent measurement for $d/a=2.55(2)$ using the bound to free spectrum at 653.55 G yields $\eb/\er=5.25(2)$, showing that bound to bound transitions correctly indicate binding energies.

The BCS side of the 2D BEC-BCS crossover is reached in Fig.~\ref{fig:bb}(c) by increasing the number of atoms to increase $E_F$, and increasing the magnetic field to reach a lower binding energy. In Fig.~\ref{fig:bb}(c) the central Fermi energy is $h\cdot 43(6)$ kHz and $T/T_F=0.5(2)$. The magnetic field is set to 834.4(1) G, where $d/a = -3.06(1)$, and the final state interactions between \ket{1} and \ket{2} are resonant, with $d/a'=-0.01(3)$. The lattice depth is $\depth/\er=26.4(3)$. Thus we know $\ebf=0.232(16)\hbar\omega_z = 2.4(2)\er$ at this lattice depth. From the bound to bound transition in Fig.~\ref{fig:bb}(c) we can then directly determine the binding energy of \ket{1}-\ket{3} fermion pairs to be $\eb/\er=0.9(2)$. The theoretical prediction~\cite{orso05form} for two-body binding gives $\eb/\er=0.82(1)$. The measured binding energy gives a many-body interaction parameter of $\lnkfa=0.6(1)$, on the BCS side but within the strongly interacting regime where one expects many-body effects beyond mean-field BEC-BCS theory~\cite{bert11crossover,pietila11pairing}. It is therefore interesting that the measured binding energy is close to the expected two-body binding energy to much better than the Fermi energy, as predicted by mean-field theory~\cite{rand89}.

In conclusion, we have measured the binding energy of fermion pairs along the crossover from 3D to 2D in a one-dimensional optical lattice. Measurements were performed at several lattice depths and scattering lengths, allowing quantitative comparison with theoretical predictions. Considering the fact that the gas is a strongly interacting many-body system, the close agreement with two-body theory is surprising, especially in the strong-coupling regime. While mean-field BEC-BCS theory in 2D predicts this behavior~\cite{rand89}, it misses other important features of the many-body system, most strikingly the interaction between fermion pairs~\cite{zhang08bcs}. Superfluidity in a one-dimensional lattice will be an exciting topic for future studies. Stacks of weakly coupled, superfluid 2D layers would constitute a basic model of the geometry found in high-temperature superconductors.

\begin{acknowledgments}
The authors would like to thank G. Orso for providing his code to calculate binding energies and M. K\"ohl and W. Zwerger for stimulating discussions. This work was supported by the NSF, AFOSR-MURI, ARO-MURI, ONR, DARPA YFA, a grant from the Army Research Office with funding from the DARPA OLE program, the David and Lucile Packard Foundation and the Alfred P. Sloan Foundation.
\end{acknowledgments}

\begin{thebibliography}{30}%
\makeatletter
\providecommand \@ifxundefined [1]{%
 \@ifx{#1\undefined}
}%
\providecommand \@ifnum [1]{%
 \ifnum #1\expandafter \@firstoftwo
 \else \expandafter \@secondoftwo
 \fi
}%
\providecommand \@ifx [1]{%
 \ifx #1\expandafter \@firstoftwo
 \else \expandafter \@secondoftwo
 \fi
}%
\providecommand \natexlab [1]{#1}%
\providecommand \enquote  [1]{``#1''}%
\providecommand \bibnamefont  [1]{#1}%
\providecommand \bibfnamefont [1]{#1}%
\providecommand \citenamefont [1]{#1}%
\providecommand \href@noop [0]{\@secondoftwo}%
\providecommand \href [0]{\begingroup \@sanitize@url \@href}%
\providecommand \@href[1]{\@@startlink{#1}\@@href}%
\providecommand \@@href[1]{\endgroup#1\@@endlink}%
\providecommand \@sanitize@url [0]{\catcode `\\12\catcode `\$12\catcode
  `\&12\catcode `\#12\catcode `\^12\catcode `\_12\catcode `\%12\relax}%
\providecommand \@@startlink[1]{}%
\providecommand \@@endlink[0]{}%
\providecommand \url  [0]{\begingroup\@sanitize@url \@url }%
\providecommand \@url [1]{\endgroup\@href {#1}{\urlprefix }}%
\providecommand \urlprefix  [0]{URL }%
\providecommand \Eprint [0]{\href }%
\providecommand \doibase [0]{http://dx.doi.org/}%
\providecommand \selectlanguage [0]{\@gobble}%
\providecommand \bibinfo  [0]{\@secondoftwo}%
\providecommand \bibfield  [0]{\@secondoftwo}%
\providecommand \translation [1]{[#1]}%
\providecommand \BibitemOpen [0]{}%
\providecommand \bibitemStop [0]{}%
\providecommand \bibitemNoStop [0]{.\EOS\space}%
\providecommand \EOS [0]{\spacefactor3000\relax}%
\providecommand \BibitemShut  [1]{\csname bibitem#1\endcsname}%
\let\auto@bib@innerbib\@empty
\bibitem [{\citenamefont {Tinkham}(2004)}]{tink04scLD}%
  \BibitemOpen
  \bibfield  {author} {\bibinfo {author} {\bibfnamefont {M.}~\bibnamefont
  {Tinkham}},\ }\href@noop {} {\emph {\bibinfo {title} {Introduction to
  Superconductivity}}},\ \bibinfo {edition} {2nd}\ ed.\ (\bibinfo  {publisher}
  {Dover},\ \bibinfo {address} {Mineola, New York},\ \bibinfo {year} {2004})\
  pp.\ \bibinfo {pages} {318--326}\BibitemShut {NoStop}%
\bibitem [{\citenamefont {Lang}(1996)}]{lang96quasi}%
  \BibitemOpen
  \bibfield  {author} {\bibinfo {author} {\bibfnamefont {M.}~\bibnamefont
  {Lang}},\ }\href@noop {} {\bibfield  {journal} {\bibinfo  {journal}
  {Supercond. Rev.}\ }\textbf {\bibinfo {volume} {2}},\ \bibinfo {pages} {1}
  (\bibinfo {year} {1996})}\BibitemShut {NoStop}%
\bibitem [{\citenamefont {Inguscio}\ \emph {et~al.}(2008)\citenamefont
  {Inguscio}, \citenamefont {Ketterle},\ and\ \citenamefont
  {Salomon}}]{ingu08varenna}%
  \BibitemOpen
  \bibinfo {editor} {\bibfnamefont {M.}~\bibnamefont {Inguscio}}, \bibinfo
  {editor} {\bibfnamefont {W.}~\bibnamefont {Ketterle}}, \ and\ \bibinfo
  {editor} {\bibfnamefont {C.}~\bibnamefont {Salomon}},\ eds.,\ \href@noop {}
  {\emph {\bibinfo {title} {Ultracold Fermi Gases}}},\ Proceedings of the
  International School of Physics "Enrico Fermi", Course CLXIV, Varenna, 20 -
  30 June 2006\ (\bibinfo  {publisher} {IOS Press, Amsterdam},\ \bibinfo {year}
  {2008})\BibitemShut {NoStop}%
\bibitem [{\citenamefont {Giorgini}\ \emph {et~al.}(2008)\citenamefont
  {Giorgini}, \citenamefont {Pitaevskii},\ and\ \citenamefont
  {Stringari}}]{giorgini08theory}%
  \BibitemOpen
  \bibfield  {author} {\bibinfo {author} {\bibfnamefont {S.}~\bibnamefont
  {Giorgini}}, \bibinfo {author} {\bibfnamefont {L.~P.}\ \bibnamefont
  {Pitaevskii}}, \ and\ \bibinfo {author} {\bibfnamefont {S.}~\bibnamefont
  {Stringari}},\ }\href {\doibase 10.1103/RevModPhys.80.1215} {\bibfield
  {journal} {\bibinfo  {journal} {Rev. Mod. Phys.}\ }\textbf {\bibinfo {volume}
  {80}},\ \bibinfo {pages} {1215} (\bibinfo {year} {2008})}\BibitemShut
  {NoStop}%
\bibitem [{\citenamefont {G\"unter}\ \emph {et~al.}(2005)\citenamefont
  {G\"unter}, \citenamefont {St\"oferle}, \citenamefont {Moritz}, \citenamefont
  {K\"ohl},\ and\ \citenamefont {Esslinger}}]{gunter05pwave}%
  \BibitemOpen
  \bibfield  {author} {\bibinfo {author} {\bibfnamefont {K.}~\bibnamefont
  {G\"unter}}, \bibinfo {author} {\bibfnamefont {T.}~\bibnamefont
  {St\"oferle}}, \bibinfo {author} {\bibfnamefont {H.}~\bibnamefont {Moritz}},
  \bibinfo {author} {\bibfnamefont {M.}~\bibnamefont {K\"ohl}}, \ and\ \bibinfo
  {author} {\bibfnamefont {T.}~\bibnamefont {Esslinger}},\ }\href {\doibase
  10.1103/PhysRevLett.95.230401} {\bibfield  {journal} {\bibinfo  {journal}
  {Phys. Rev. Lett.}\ }\textbf {\bibinfo {volume} {95}},\ \bibinfo {pages}
  {230401} (\bibinfo {year} {2005})}\BibitemShut {NoStop}%
\bibitem [{\citenamefont {Du}\ \emph {et~al.}(2009)\citenamefont {Du},
  \citenamefont {Zhang},\ and\ \citenamefont {Thomas}}]{du09inelastic}%
  \BibitemOpen
  \bibfield  {author} {\bibinfo {author} {\bibfnamefont {X.}~\bibnamefont
  {Du}}, \bibinfo {author} {\bibfnamefont {Y.}~\bibnamefont {Zhang}}, \ and\
  \bibinfo {author} {\bibfnamefont {J.~E.}\ \bibnamefont {Thomas}},\ }\href
  {\doibase 10.1103/PhysRevLett.102.250402} {\bibfield  {journal} {\bibinfo
  {journal} {Phys. Rev. Lett.}\ }\textbf {\bibinfo {volume} {102}},\ \bibinfo
  {pages} {250402} (\bibinfo {year} {2009})}\BibitemShut {NoStop}%
\bibitem [{\citenamefont {Martiyanov}\ \emph {et~al.}(2010)\citenamefont
  {Martiyanov}, \citenamefont {Makhalov},\ and\ \citenamefont
  {Turlapov}}]{martiyanov10observation}%
  \BibitemOpen
  \bibfield  {author} {\bibinfo {author} {\bibfnamefont {K.}~\bibnamefont
  {Martiyanov}}, \bibinfo {author} {\bibfnamefont {V.}~\bibnamefont
  {Makhalov}}, \ and\ \bibinfo {author} {\bibfnamefont {A.}~\bibnamefont
  {Turlapov}},\ }\href {\doibase 10.1103/PhysRevLett.105.030404} {\bibfield
  {journal} {\bibinfo  {journal} {Phys. Rev. Lett.}\ }\textbf {\bibinfo
  {volume} {105}},\ \bibinfo {pages} {030404} (\bibinfo {year}
  {2010})}\BibitemShut {NoStop}%
\bibitem [{\citenamefont {Fr\"ohlich}\ \emph {et~al.}(2011)\citenamefont
  {Fr\"ohlich}, \citenamefont {Feld}, \citenamefont {Vogt}, \citenamefont
  {Koschorreck}, \citenamefont {Zwerger},\ and\ \citenamefont
  {K\"ohl}}]{frohlich11radio}%
  \BibitemOpen
  \bibfield  {author} {\bibinfo {author} {\bibfnamefont {B.}~\bibnamefont
  {Fr\"ohlich}}, \bibinfo {author} {\bibfnamefont {M.}~\bibnamefont {Feld}},
  \bibinfo {author} {\bibfnamefont {E.}~\bibnamefont {Vogt}}, \bibinfo {author}
  {\bibfnamefont {M.}~\bibnamefont {Koschorreck}}, \bibinfo {author}
  {\bibfnamefont {W.}~\bibnamefont {Zwerger}}, \ and\ \bibinfo {author}
  {\bibfnamefont {M.}~\bibnamefont {K\"ohl}},\ }\href {\doibase
  10.1103/PhysRevLett.106.105301} {\bibfield  {journal} {\bibinfo  {journal}
  {Phys. Rev. Lett.}\ }\textbf {\bibinfo {volume} {106}},\ \bibinfo {pages}
  {105301} (\bibinfo {year} {2011})}\BibitemShut {NoStop}%
\bibitem [{\citenamefont {Dyke}\ \emph {et~al.}(2011)\citenamefont {Dyke},
  \citenamefont {Kuhnle}, \citenamefont {Whitlock}, \citenamefont {Hu},
  \citenamefont {Mark}, \citenamefont {Hoinka}, \citenamefont {Lingham},
  \citenamefont {Hannaford},\ and\ \citenamefont {Vale}}]{dyke11crossover}%
  \BibitemOpen
  \bibfield  {author} {\bibinfo {author} {\bibfnamefont {P.}~\bibnamefont
  {Dyke}}, \bibinfo {author} {\bibfnamefont {E.~D.}\ \bibnamefont {Kuhnle}},
  \bibinfo {author} {\bibfnamefont {S.}~\bibnamefont {Whitlock}}, \bibinfo
  {author} {\bibfnamefont {H.}~\bibnamefont {Hu}}, \bibinfo {author}
  {\bibfnamefont {M.}~\bibnamefont {Mark}}, \bibinfo {author} {\bibfnamefont
  {S.}~\bibnamefont {Hoinka}}, \bibinfo {author} {\bibfnamefont
  {M.}~\bibnamefont {Lingham}}, \bibinfo {author} {\bibfnamefont
  {P.}~\bibnamefont {Hannaford}}, \ and\ \bibinfo {author} {\bibfnamefont
  {C.~J.}\ \bibnamefont {Vale}},\ }\href {\doibase
  10.1103/PhysRevLett.106.105304} {\bibfield  {journal} {\bibinfo  {journal}
  {Phys. Rev. Lett.}\ }\textbf {\bibinfo {volume} {106}},\ \bibinfo {pages}
  {105304} (\bibinfo {year} {2011})}\BibitemShut {NoStop}%
\bibitem [{\citenamefont {Orso}\ and\ \citenamefont
  {Shlyapnikov}(2005)}]{orso05super}%
  \BibitemOpen
  \bibfield  {author} {\bibinfo {author} {\bibfnamefont {G.}~\bibnamefont
  {Orso}}\ and\ \bibinfo {author} {\bibfnamefont {G.~V.}\ \bibnamefont
  {Shlyapnikov}},\ }\href {\doibase 10.1103/PhysRevLett.95.260402} {\bibfield
  {journal} {\bibinfo  {journal} {Phys. Rev. Lett.}\ }\textbf {\bibinfo
  {volume} {95}},\ \bibinfo {pages} {260402} (\bibinfo {year}
  {2005})}\BibitemShut {NoStop}%
\bibitem [{\citenamefont {Orso}\ \emph {et~al.}(2005)\citenamefont {Orso},
  \citenamefont {Pitaevskii}, \citenamefont {Stringari},\ and\ \citenamefont
  {Wouters}}]{orso05form}%
  \BibitemOpen
  \bibfield  {author} {\bibinfo {author} {\bibfnamefont {G.}~\bibnamefont
  {Orso}}, \bibinfo {author} {\bibfnamefont {L.~P.}\ \bibnamefont
  {Pitaevskii}}, \bibinfo {author} {\bibfnamefont {S.}~\bibnamefont
  {Stringari}}, \ and\ \bibinfo {author} {\bibfnamefont {M.}~\bibnamefont
  {Wouters}},\ }\href {\doibase 10.1103/PhysRevLett.95.060402} {\bibfield
  {journal} {\bibinfo  {journal} {Phys. Rev. Lett.}\ }\textbf {\bibinfo
  {volume} {95}},\ \bibinfo {pages} {060402} (\bibinfo {year}
  {2005})}\BibitemShut {NoStop}%
\bibitem [{\citenamefont {Salasnich}(2007)}]{salasnich07condensate}%
  \BibitemOpen
  \bibfield  {author} {\bibinfo {author} {\bibfnamefont {L.}~\bibnamefont
  {Salasnich}},\ }\href {\doibase 10.1103/PhysRevA.76.015601} {\bibfield
  {journal} {\bibinfo  {journal} {Phys. Rev. A}\ }\textbf {\bibinfo {volume}
  {76}},\ \bibinfo {pages} {015601} (\bibinfo {year} {2007})}\BibitemShut
  {NoStop}%
\bibitem [{\citenamefont {Zhang}\ \emph
  {et~al.}(2008{\natexlab{a}})\citenamefont {Zhang}, \citenamefont {Lin},\ and\
  \citenamefont {Duan}}]{zhang08bcs}%
  \BibitemOpen
  \bibfield  {author} {\bibinfo {author} {\bibfnamefont {W.}~\bibnamefont
  {Zhang}}, \bibinfo {author} {\bibfnamefont {G.-D.}\ \bibnamefont {Lin}}, \
  and\ \bibinfo {author} {\bibfnamefont {L.-M.}\ \bibnamefont {Duan}},\ }\href
  {\doibase 10.1103/PhysRevA.77.063613} {\bibfield  {journal} {\bibinfo
  {journal} {Phys. Rev. A}\ }\textbf {\bibinfo {volume} {77}},\ \bibinfo
  {pages} {063613} (\bibinfo {year} {2008}{\natexlab{a}})}\BibitemShut
  {NoStop}%
\bibitem [{\citenamefont {Ruggiero}\ \emph {et~al.}(1980)\citenamefont
  {Ruggiero}, \citenamefont {Barbee},\ and\ \citenamefont
  {Beasley}}]{rugg80quasi2D}%
  \BibitemOpen
  \bibfield  {author} {\bibinfo {author} {\bibfnamefont {S.~T.}\ \bibnamefont
  {Ruggiero}}, \bibinfo {author} {\bibfnamefont {T.~W.}\ \bibnamefont
  {Barbee}}, \ and\ \bibinfo {author} {\bibfnamefont {M.~R.}\ \bibnamefont
  {Beasley}},\ }\href {\doibase 10.1103/PhysRevLett.45.1299} {\bibfield
  {journal} {\bibinfo  {journal} {Phys. Rev. Lett.}\ }\textbf {\bibinfo
  {volume} {45}},\ \bibinfo {pages} {1299} (\bibinfo {year}
  {1980})}\BibitemShut {NoStop}%
\bibitem [{\citenamefont {Kosterlitz}\ and\ \citenamefont
  {Thouless}(1972)}]{kosterlitz72long}%
  \BibitemOpen
  \bibfield  {author} {\bibinfo {author} {\bibfnamefont {J.~M.}\ \bibnamefont
  {Kosterlitz}}\ and\ \bibinfo {author} {\bibfnamefont {D.}~\bibnamefont
  {Thouless}},\ }\href@noop {} {\bibfield  {journal} {\bibinfo  {journal} {J.
  Phys. C.}\ }\textbf {\bibinfo {volume} {5}},\ \bibinfo {pages} {L124}
  (\bibinfo {year} {1972})}\BibitemShut {NoStop}%
\bibitem [{\citenamefont {Petrov}\ \emph {et~al.}(2003)\citenamefont {Petrov},
  \citenamefont {Baranov},\ and\ \citenamefont
  {Shlyapnikov}}]{petrov03superfluid}%
  \BibitemOpen
  \bibfield  {author} {\bibinfo {author} {\bibfnamefont {D.~S.}\ \bibnamefont
  {Petrov}}, \bibinfo {author} {\bibfnamefont {M.~A.}\ \bibnamefont {Baranov}},
  \ and\ \bibinfo {author} {\bibfnamefont {G.~V.}\ \bibnamefont
  {Shlyapnikov}},\ }\href {\doibase 10.1103/PhysRevA.67.031601} {\bibfield
  {journal} {\bibinfo  {journal} {Phys. Rev. A}\ }\textbf {\bibinfo {volume}
  {67}},\ \bibinfo {pages} {031601} (\bibinfo {year} {2003})}\BibitemShut
  {NoStop}%
\bibitem [{\citenamefont {Zhang}\ \emph
  {et~al.}(2008{\natexlab{b}})\citenamefont {Zhang}, \citenamefont {Lin},\ and\
  \citenamefont {Duan}}]{zhang08berez}%
  \BibitemOpen
  \bibfield  {author} {\bibinfo {author} {\bibfnamefont {W.}~\bibnamefont
  {Zhang}}, \bibinfo {author} {\bibfnamefont {G.-D.}\ \bibnamefont {Lin}}, \
  and\ \bibinfo {author} {\bibfnamefont {L.-M.}\ \bibnamefont {Duan}},\ }\href
  {\doibase 10.1103/PhysRevA.78.043617} {\bibfield  {journal} {\bibinfo
  {journal} {Phys. Rev. A}\ }\textbf {\bibinfo {volume} {78}},\ \bibinfo
  {pages} {043617} (\bibinfo {year} {2008}{\natexlab{b}})}\BibitemShut
  {NoStop}%
\bibitem [{\citenamefont {Iskin}\ and\ \citenamefont
  {de~Melo}(2009)}]{iskin09evolution}%
  \BibitemOpen
  \bibfield  {author} {\bibinfo {author} {\bibfnamefont {M.}~\bibnamefont
  {Iskin}}\ and\ \bibinfo {author} {\bibfnamefont {C.~A. R.~S.}\ \bibnamefont
  {de~Melo}},\ }\href {\doibase 10.1103/PhysRevLett.103.165301} {\bibfield
  {journal} {\bibinfo  {journal} {Phys. Rev. Lett.}\ }\textbf {\bibinfo
  {volume} {103}},\ \bibinfo {pages} {165301} (\bibinfo {year}
  {2009})}\BibitemShut {NoStop}%
\bibitem [{\citenamefont {Cooper}(1956)}]{coop56boun}%
  \BibitemOpen
  \bibfield  {author} {\bibinfo {author} {\bibfnamefont {L.~N.}\ \bibnamefont
  {Cooper}},\ }\href {\doibase 10.1103/PhysRev.104.1189} {\bibfield  {journal}
  {\bibinfo  {journal} {Phys. Rev.}\ }\textbf {\bibinfo {volume} {104}},\
  \bibinfo {pages} {1189} (\bibinfo {year} {1956})}\BibitemShut {NoStop}%
\bibitem [{\citenamefont {Randeria}\ \emph {et~al.}(1989)\citenamefont
  {Randeria}, \citenamefont {Duan},\ and\ \citenamefont {Shieh}}]{rand89}%
  \BibitemOpen
  \bibfield  {author} {\bibinfo {author} {\bibfnamefont {M.}~\bibnamefont
  {Randeria}}, \bibinfo {author} {\bibfnamefont {J.-M.}\ \bibnamefont {Duan}},
  \ and\ \bibinfo {author} {\bibfnamefont {L.}~\bibnamefont {Shieh}},\
  }\href@noop {} {\bibfield  {journal} {\bibinfo  {journal} {Phys. Rev. Lett.}\
  }\textbf {\bibinfo {volume} {62}},\ \bibinfo {pages} {981} (\bibinfo {year}
  {1989})}\BibitemShut {NoStop}%
\bibitem [{\citenamefont {Ketterle}\ and\ \citenamefont
  {Zwierlein}(2008)}]{kett08maki}%
  \BibitemOpen
  \bibfield  {author} {\bibinfo {author} {\bibfnamefont {W.}~\bibnamefont
  {Ketterle}}\ and\ \bibinfo {author} {\bibfnamefont {M.}~\bibnamefont
  {Zwierlein}},\ }\href {\doibase 10.1393/ncr/i2008-10033-1} {\bibfield
  {journal} {\bibinfo  {journal} {La Rivista del Nuovo Cimento}\ }\textbf
  {\bibinfo {volume} {31}},\ \bibinfo {pages} {247} (\bibinfo {year}
  {2008})}\BibitemShut {NoStop}%
\bibitem [{\citenamefont {Petrov}\ and\ \citenamefont
  {Shlyapnikov}(2001)}]{petr01inte}%
  \BibitemOpen
  \bibfield  {author} {\bibinfo {author} {\bibfnamefont {D.~S.}\ \bibnamefont
  {Petrov}}\ and\ \bibinfo {author} {\bibfnamefont {G.~V.}\ \bibnamefont
  {Shlyapnikov}},\ }\href {\doibase 10.1103/PhysRevA.64.012706} {\bibfield
  {journal} {\bibinfo  {journal} {Phys. Rev. A}\ }\textbf {\bibinfo {volume}
  {64}},\ \bibinfo {pages} {012706} (\bibinfo {year} {2001})}\BibitemShut
  {NoStop}%
\bibitem [{\citenamefont {Bloch}\ \emph {et~al.}(2008)\citenamefont {Bloch},
  \citenamefont {Dalibard},\ and\ \citenamefont {Zwerger}}]{bloc08many}%
  \BibitemOpen
  \bibfield  {author} {\bibinfo {author} {\bibfnamefont {I.}~\bibnamefont
  {Bloch}}, \bibinfo {author} {\bibfnamefont {J.}~\bibnamefont {Dalibard}}, \
  and\ \bibinfo {author} {\bibfnamefont {W.}~\bibnamefont {Zwerger}},\ }\href
  {\doibase 10.1103/RevModPhys.80.885} {\bibfield  {journal} {\bibinfo
  {journal} {Rev. Mod. Phys.}\ }\textbf {\bibinfo {volume} {80}},\ \bibinfo
  {pages} {885} (\bibinfo {year} {2008})}\BibitemShut {NoStop}%
\bibitem [{\citenamefont {Schunck}\ \emph {et~al.}(2008)\citenamefont
  {Schunck}, \citenamefont {Shin}, \citenamefont {Schirotzek},\ and\
  \citenamefont {Ketterle}}]{schu08dete}%
  \BibitemOpen
  \bibfield  {author} {\bibinfo {author} {\bibfnamefont {C.}~\bibnamefont
  {Schunck}}, \bibinfo {author} {\bibfnamefont {Y.}~\bibnamefont {Shin}},
  \bibinfo {author} {\bibfnamefont {A.}~\bibnamefont {Schirotzek}}, \ and\
  \bibinfo {author} {\bibfnamefont {W.}~\bibnamefont {Ketterle}},\ }\href
  {\doibase 10.1038/nature07176} {\bibfield  {journal} {\bibinfo  {journal}
  {Nature}\ }\textbf {\bibinfo {volume} {454}},\ \bibinfo {pages} {739}
  (\bibinfo {year} {2008})}\BibitemShut {NoStop}%
\bibitem [{\citenamefont {Bartenstein}\ \emph {et~al.}(2005)\citenamefont
  {Bartenstein}, \citenamefont {Altmeyer}, \citenamefont {Riedl}, \citenamefont
  {Geursen}, \citenamefont {Jochim}, \citenamefont {Chin}, \citenamefont
  {Denschlag}, \citenamefont {Grimm}, \citenamefont {Simoni}, \citenamefont
  {Tiesinga}, \citenamefont {Williams},\ and\ \citenamefont
  {Julienne}}]{bart05prec}%
  \BibitemOpen
  \bibfield  {author} {\bibinfo {author} {\bibfnamefont {M.}~\bibnamefont
  {Bartenstein}}, \bibinfo {author} {\bibfnamefont {A.}~\bibnamefont
  {Altmeyer}}, \bibinfo {author} {\bibfnamefont {S.}~\bibnamefont {Riedl}},
  \bibinfo {author} {\bibfnamefont {R.}~\bibnamefont {Geursen}}, \bibinfo
  {author} {\bibfnamefont {S.}~\bibnamefont {Jochim}}, \bibinfo {author}
  {\bibfnamefont {C.}~\bibnamefont {Chin}}, \bibinfo {author} {\bibfnamefont
  {J.~H.}\ \bibnamefont {Denschlag}}, \bibinfo {author} {\bibfnamefont
  {R.}~\bibnamefont {Grimm}}, \bibinfo {author} {\bibfnamefont
  {A.}~\bibnamefont {Simoni}}, \bibinfo {author} {\bibfnamefont
  {E.}~\bibnamefont {Tiesinga}}, \bibinfo {author} {\bibfnamefont {C.~J.}\
  \bibnamefont {Williams}}, \ and\ \bibinfo {author} {\bibfnamefont {P.~S.}\
  \bibnamefont {Julienne}},\ }\href {\doibase 10.1103/PhysRevLett.94.103201}
  {\bibfield  {journal} {\bibinfo  {journal} {Phys. Rev. Lett.}\ }\textbf
  {\bibinfo {volume} {94}},\ \bibinfo {pages} {103201} (\bibinfo {year}
  {2005})}\BibitemShut {NoStop}%
\bibitem [{\citenamefont {Pietila}\ \emph {et~al.}(2011)\citenamefont
  {Pietila}, \citenamefont {Pekker}, \citenamefont {Nishida},\ and\
  \citenamefont {Demler}}]{pietila11pairing}%
  \BibitemOpen
  \bibfield  {author} {\bibinfo {author} {\bibfnamefont {V.}~\bibnamefont
  {Pietila}}, \bibinfo {author} {\bibfnamefont {D.}~\bibnamefont {Pekker}},
  \bibinfo {author} {\bibfnamefont {Y.}~\bibnamefont {Nishida}}, \ and\
  \bibinfo {author} {\bibfnamefont {E.}~\bibnamefont {Demler}},\ }\href@noop {}
  {\bibfield  {journal} {\bibinfo  {journal} {arxiv:cond-mat/1110.0494}\ }
  (\bibinfo {year} {2011})}\BibitemShut {NoStop}%
\bibitem [{\citenamefont {Chin}\ and\ \citenamefont
  {Julienne}(2005)}]{chin05radi}%
  \BibitemOpen
  \bibfield  {author} {\bibinfo {author} {\bibfnamefont {C.}~\bibnamefont
  {Chin}}\ and\ \bibinfo {author} {\bibfnamefont {P.~S.}\ \bibnamefont
  {Julienne}},\ }\href {\doibase 10.1103/PhysRevA.71.012713} {\bibfield
  {journal} {\bibinfo  {journal} {Phys. Rev. A}\ }\textbf {\bibinfo {volume}
  {71}},\ \bibinfo {pages} {012713} (\bibinfo {year} {2005})}\BibitemShut
  {NoStop}%
\bibitem [{\citenamefont {Langmack}\ \emph {et~al.}(2011)\citenamefont
  {Langmack}, \citenamefont {Barth}, \citenamefont {Zwerger},\ and\
  \citenamefont {Braaten}}]{lang11clock2d}%
  \BibitemOpen
  \bibfield  {author} {\bibinfo {author} {\bibfnamefont {C.}~\bibnamefont
  {Langmack}}, \bibinfo {author} {\bibfnamefont {M.}~\bibnamefont {Barth}},
  \bibinfo {author} {\bibfnamefont {W.}~\bibnamefont {Zwerger}}, \ and\
  \bibinfo {author} {\bibfnamefont {E.}~\bibnamefont {Braaten}},\ }\href
  {http://arxiv.org/abs/1111.0999} {\bibfield  {journal} {\bibinfo  {journal}
  {preprint arXiv:1111.0999}\ } (\bibinfo {year} {2011})}\BibitemShut {NoStop}%
\bibitem [{\citenamefont {Bertaina}\ and\ \citenamefont
  {Giorgini}(2011)}]{bert11crossover}%
  \BibitemOpen
  \bibfield  {author} {\bibinfo {author} {\bibfnamefont {G.}~\bibnamefont
  {Bertaina}}\ and\ \bibinfo {author} {\bibfnamefont {S.}~\bibnamefont
  {Giorgini}},\ }\href {\doibase 10.1103/PhysRevLett.106.110403} {\bibfield
  {journal} {\bibinfo  {journal} {Phys. Rev. Lett.}\ }\textbf {\bibinfo
  {volume} {106}},\ \bibinfo {pages} {110403} (\bibinfo {year}
  {2011})}\BibitemShut {NoStop}%
\bibitem [{\citenamefont {Schirotzek}\ \emph {et~al.}(2008)\citenamefont
  {Schirotzek}, \citenamefont {Shin}, \citenamefont {Schunck},\ and\
  \citenamefont {Ketterle}}]{schi08dete}%
  \BibitemOpen
  \bibfield  {author} {\bibinfo {author} {\bibfnamefont {A.}~\bibnamefont
  {Schirotzek}}, \bibinfo {author} {\bibfnamefont {Y.-i.}\ \bibnamefont
  {Shin}}, \bibinfo {author} {\bibfnamefont {C.~H.}\ \bibnamefont {Schunck}}, \
  and\ \bibinfo {author} {\bibfnamefont {W.}~\bibnamefont {Ketterle}},\ }\href
  {\doibase 10.1103/PhysRevLett.101.140403} {\bibfield  {journal} {\bibinfo
  {journal} {Phys. Rev. Lett.}\ }\textbf {\bibinfo {volume} {101}},\ \bibinfo
  {pages} {140403} (\bibinfo {year} {2008})}\BibitemShut {NoStop}%
\end{thebibliography}%
%

\end{document}